\documentclass[twocolumn]{emulateapj}
\usepackage{epsf}
\usepackage{amsmath}
\usepackage{multirow}
\usepackage{rotating}
\usepackage{subfigure}
\usepackage{multirow}
\usepackage{color}

\begin{document}

\shortauthors{A. L. King et al.}
\shorttitle{4U~1630$-$472}

\title{The Disk Wind in the Rapidly Spinning Stellar-Mass Black Hole 4U~1630$-$472 Observed with \textit{NuSTAR}}

\author{Ashley~L.~King\altaffilmark{1},
	Dominic J. Walton\altaffilmark{2},
	Jon M. Miller\altaffilmark{1},
	Didier Barret\altaffilmark{3,4},
	Steven E. Boggs\altaffilmark{5},
	Finn E. Christensen\altaffilmark{6},
	William W. Craig\altaffilmark{5,7},
	Andy C. Fabian\altaffilmark{8},
	Felix F\"urst\altaffilmark{2},
	Charles J. Hailey\altaffilmark{9},
	Fiona A. Harrison\altaffilmark{2},
	Roman Krivonos\altaffilmark{5},
	Kaya Mori\altaffilmark{9},
	Lorenzo Natalucci\altaffilmark{10},
	Daniel Stern\altaffilmark{11},
	John A. Tomsick\altaffilmark{5},
	William W. Zhang\altaffilmark{12}}
        
\altaffiltext{1}{Department of Astronomy, University of Michigan, 500
Church Street, Ann Arbor, MI 48109-1042, ashking@umich.edu}
\altaffiltext{2}{Cahill Center for Astronomy and Astrophysics, California Institute of Technology, Pasadena, CA 91125, USA}
\altaffiltext{3}{UniversitŽ de Toulouse; UPS-OMP; IRAP; Toulouse, France}
\altaffiltext{4}{CNRS; Institut de Recherche en Astrophysique et PlanŽtologie; 9 Av. colonel Roche, BP 44346, F-31028 Toulouse cedex 4, France}
\altaffiltext{5}{Space Sciences Laboratory, 7 Gauss Way, University of California, Berkeley, CA 94720-7450, USA}
\altaffiltext{6}{DTU Space, National Space Institute, Technical University of Denmark, Elektrovej 327, DK-2800 Lyngby, Denmark}
\altaffiltext{7}{Lawrence Livermore National Laboratory, Livermore, CA 94550, USA}
\altaffiltext{8}{Institute of Astronomy, University of Cambridge, Madingley Road, Cambridge CB3 0HA, UK}
\altaffiltext{9}{Columbia Astrophysics Laboratory, Columbia University, New York, NY 10027, USA}
\altaffiltext{10}{Istituto Nazionale di Astrofisica, INAF-IAPS, via del Fosso del Cavaliere, 00133 Roma, Italy}
\altaffiltext{11}{Jet Propulsion Laboratory, California Institute of
Technology, 4800 Oak Grove Drive, Mail Stop 169-221, Pasadena, CA
91109}
\altaffiltext{12}{NASA Goddard Space Flight Center, Greenbelt, MD 20771}

\label{firstpage}
\begin{abstract}
We present an analysis of a short {\it NuSTAR} observation of the
stellar-mass black hole and low-mass X-ray binary 4U~1630$-$472.
Reflection from the inner accretion disk is clearly detected for the
first time in this source, owing to the sensitivity of {\it NuSTAR}.
With fits to the reflection spectrum, we find evidence for a rapidly
spinning black hole, $a_{*}=0.985^{+0.005}_{-0.014}$~($1\sigma$
statistical errors). However, archival data show that the source has relatively low radio luminosity. Recently claimed relationships between jet power and black hole spin would predict either a lower spin or a higher peak radio luminosity.  We also report the clear detection of an
absorption feature at $7.03\pm0.03$~keV, likely signaling a disk wind. If this line arises in dense, moderately
ionized gas ($\log\xi=3.6^{+0.2}_{-0.3}$) and is dominated by He-like Fe
XXV, the wind has a velocity of $v/c=0.043^{+0.002}_{-0.007}$~(12900$^{+600}_{-2100}$~km~s$^{-1}$).  If the line is
instead associated with a more highly ionized gas ($\log\xi=6.1^{+0.7}_{-0.6}$), and is dominated by Fe XXVI, evidence of a blue-shift is only marginal, after taking systematic errors into account. Our analysis suggests the ionized wind may be launched within 200--1100 Rg, and may be magnetically driven.
\end{abstract}
 

\section{Introduction}

Direct observations of the most powerful outflows are key to advancing our understanding of  the mechanisms that launch and drive them, and also how these winds influence galactic environments. Though active galactic nuclei ultimately have the most powerful outflows in the Universe \citep[e.g.,][]{Allen06}, stellar-mass black holes operate on time-scales that are easier to study \citep[e.g.][]{King12}. Studies of such sources can further our understanding of these powerful outflows not only on stellar scales but on galactic scales as well. 

In stellar-mass black holes, winds and jets appear to be anti-correlated and associated with particular X-ray spectral states \citep{Miller06a,Miller08,Neilsen09,Ponti12}. Highly ionized X-ray winds are found in luminous and spectrally soft states \citep[see ][for a theoretical exploration of the thermodynamical stability of these winds as a function of spectral state]{Chakravorty13}. Conversely, relativistic jets, denoted by radio emission, are found in the spectrally hard states. It is not yet fully understood how they are related, e.g. whether one outflow quenches the other. However, jets and winds do appear to be regulated in a similar fashion and possibly by the same mechanism \citep{King13a}. In fact, in terms of their energetics, the highest velocity outflows resemble jets more so than they resemble slower velocity winds. Therefore, by examining winds in X-ray binaries, disks, winds, and jets can be studied, with implications across the black hole mass scale.

4U~1630$-$472 is a recurrent black hole X-ray binary: outbursts have been observed dating back to 1969 \citep[see, e.g.,][]{Kuulkers97a}.  It lies close to the Galactic plane, behind a large neutral column \citep[$N_{H}\simeq10^{23}~\rm{cm}^{-2}$; e.g.,][]{Parmar95}.   This has prevented optical studies that could constrain the mass of the primary.  

The evidence that 4U~1630$-$472 harbors a black hole is circumstantial, but compelling.  Its X-ray spectral and timing properties are broadly typical of black hole systems \citep[e.g.,][]{Barret96,Abe05}.   Indeed, the best evidence may come from its high--frequency timing features, which include quasi-periodic oscillations (QPO's) at frequencies up to 262.2~Hz \citep{KleinWolt04}.  

Prior observations of 4U~1630$-$472 establish that it is likely viewed at a high inclination.  In particular, flux dips have been detected with {\it RXTE} \citep{Tomsick98,Kuulkers98}, indicating a line of sight that intersects the outer accretion disk.  However, full eclipses are not observed, implying a high, but not edge-on inclination \citep[$\sim70^\circ$, comparable to GRO~J1655$-$40, which exhibits similar behavior,][]{Orosz97}.

4U~1630$-$472 has been elusive when it comes to reflection features.  Disk reflection spectra, including Fe K emission lines, are a natural consequence of any geometry in which the accretion disk is illuminated by an external source of hard X-rays.  Relativistically broadened Fe K lines are therefore a hallmark of black hole accretion \citep[for a review, see][]{Miller07}.  Yet, such features have not been clearly detected in 4U~1630$-$472 in past observations, though evidence for narrow features has been detected.  Using low--resolution {\it RXTE} data, for instance, \cite{Cui00} reported evidence of Doppler-split Fe K emission lines.  Also using {\it RXTE} data, \cite{Tomsick00} reported the detection of a neutral Fe K line in a spectrum with low flux. Finally,  \cite{Kubota07} reported detections of Fe K$\alpha$ absorption features associated with Fe XXV and Fe XXVI.  

Finally, 4U~1630$-$472 has shown unresolved radio emission indicating the production of a jet \citep{Hjellming99,Trigo13}. However, in observations by \cite{Hannikainen02}, it has also been radio quiet, even in the canonical low/hard state where jet production is thought to be prevalent.

In this Letter, we present a {\it NuSTAR} observation of 4U~1630$-$472, in which we detect both a broad iron emission line and a highly ionized outflow. We first describe the observations and analysis of these data using phenomenological models, and then proceed to  physically self-consistent models of both the reflection and absorption features.  All errors are 1$\sigma$ unless otherwise specified.

\section{ Observations}
{\it NuSTAR} \citep{Harrison13} observed 4U~1630$-$472 on UT 2013 February 21, with an exposure time of 14.7~ks (ObsID 40014009). The data were processed using the nustardas version 1.1.1 and the 2013 May 9 version of CALDB. The spectra were extracted with the standard pipeline using an extraction region of 40$''$ for both focal plane modules A and B (FPMA/B). This relatively small extraction region avoids stray light from a nearby source. The background spectra were extracted with a similar sized extraction region in an area that avoided both the incident stray light and X-rays from the source itself.  The data were binned with a minimum of 20 counts per energy bin using {\it grppha} \citep[ftools, v6.13 ][]{blackburn95}. The spectra were fit between 3$-$79 keV using Xspec, v12.8.0 \citep{Arnaud96}.

\section{Analysis}
\begin{figure*}[t]
\centering
\includegraphics[scale=.85,angle=0]{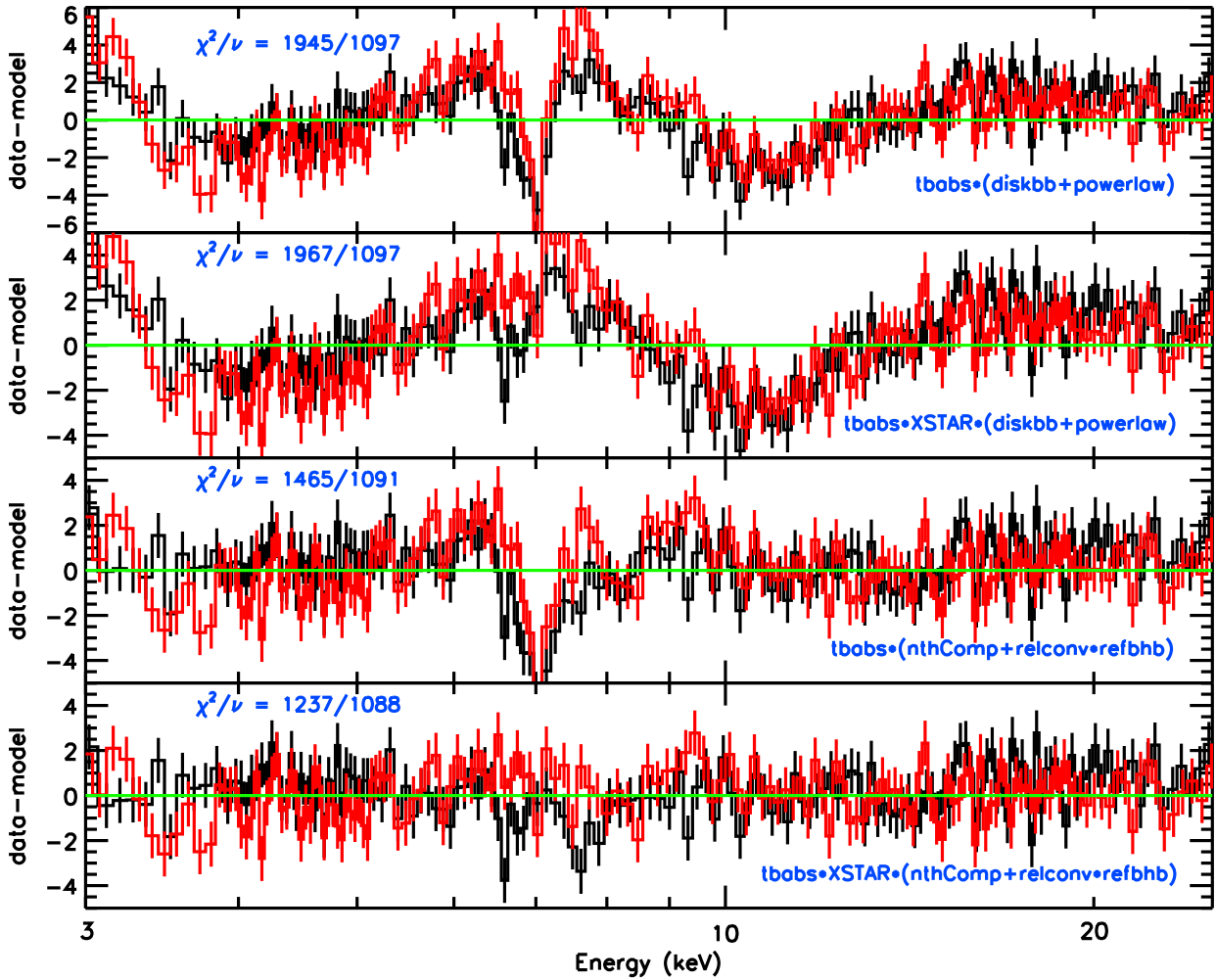}
\caption{The residuals of the data minus the model with error bars of size $1\sigma$ for several models applied to the {\it NuSTAR} spectra of 4U~1630$-$472 and rebinned for clarity. FPMA is in black and FPMB is in red. The reduced $\chi^2$ is shown in the top left of each panel.  \textit{Top panel:} the initial disk plus a powerlaw continuum model. \textit{Second panel:} the same model but with the best fit XSTAR photoionization component added; there are still clear indications of the reflection component. \textit{Third panel:} our best fit reflection model, but with no absorption component. \textit{Bottom panel:} our best fit model, including both disk reflection and ionized absorption. \label{fig:del}}
\end{figure*}

The flux and luminosity were derived using a canonical power-law and disk blackbody, modified by interstellar absorption (using TBabs; \citealt{Wilms00}). The results of this fit are given in Table 1 and the residuals are shown in Figure \ref{fig:del}. The data show residual features, indicating that the spectrum is not fully described by the fiducial model. In particular, there is a deep absorption feature at 7~keV and two broad emission components: one at low energy (E$\sim$6 keV) and one at higher energy (E$\sim$20 keV). If we replace the powerlaw model with a more physical Comptonisation continuum, e.g. {\tt nthComp} \citep[e.g.,][]{Zycki99}, linking the seed photon temperature to that of the disk, the observed residuals are unchanged, but see section 3.1.

The source was in a ``intermediate" state with an absorbed 2$-$10$\rm{keV}$ flux of 5.9$\times10^{-9}$ergs~cm$^{-2}$~s$^{-1}$, and an unabsorbed flux of 1.2$\times10^{-8}$ergs~cm$^{-2}$~s$^{-1}$ (the preliminary model was extended to 2~keV, out of the range of the {\it NuSTAR} band).  This corresponds to a 2-10~keV luminosity at 10~kpc \citep{Augusteijn01} of $L=1.4\times10^{38}$ergs~s$^{-1}$. 

To better fit the data, we now include both the contribution from relativistic disk reflection model (Section~\ref{sec:refl}), and photoionized absorption (Section~\ref{sec:abs}).

\subsection{Disk Reflection}
\label{sec:refl}
To model the broad residuals at both 6~keV and 20~keV, we utilized the reflection model, {\tt refbhb} \citep{Ross07, Reis08}, which is designed specifically for use with Galactic binaries, convolved with a relativistic blurring kernel {\tt relconv} \citep{Dauser10}, assuming solar abundances. We link the illuminating photon index ($\Gamma$) to that of the Comptonized continuum, {\tt nthComp}, which in turn has its seed photon temperature ($kT_{\rm{BB}}$) set to that of the blackbody temperature of {\tt refbhb}. Following \cite{Wilkins11} and \cite{Fabian12}, we assume a broken powerlaw for the radial emissivity profile, with the outer emissivity index set to the Newtonian approximation ($q_{\rm{out}}=3$). The additional free parameters for the reflected emission are the inner emissivity index ($q_{\rm{in}}$), the break radius ($R_{\rm{Br}}$), the disk inclination ($\theta$), density ($H_{\rm{Den}}$), the dimensionless black hole spin ($a_*\equiv\frac{cJ}{GM^2}$), the ratio of the illuminating and disk fluxes (at the disk surface; Illum/BB), and the overall reflection normalization. 

With this model, we find the Fe line in 4U~1630$-$472 to be partially ionized with $\log\xi=2.9\pm0.2$, while the system has a spin and inclination of $a_{*}=0.985^{+0.005}_{-0.014}$ and $\theta=64^{+2}_{-3}$ degrees, respectively. Although to some extent these constraints are model dependent, and the uncertainties quoted here are purely statistical, studies that have investigated a variety of modeling assumptions for other rapidly rotating black holes have generally shown the high spin values obtained to be robust (e.g. \citealt{Reis12,Walton12,Tomsick13}). Indeed, if we replace {\tt refbhb} with {\tt reflionx} and any standard accretion disk model, we obtain the same spin. Furthermore, recent MHD simulations suggest that emission from inside the innermost stable circular orbit (ISCO) should be negligible \citep{Reynolds08}, supporting the assumption that the disk truncates at the ISCO, which is fundamental to spin measurements.

\begin{figure}
\includegraphics[scale=.55,angle=0]{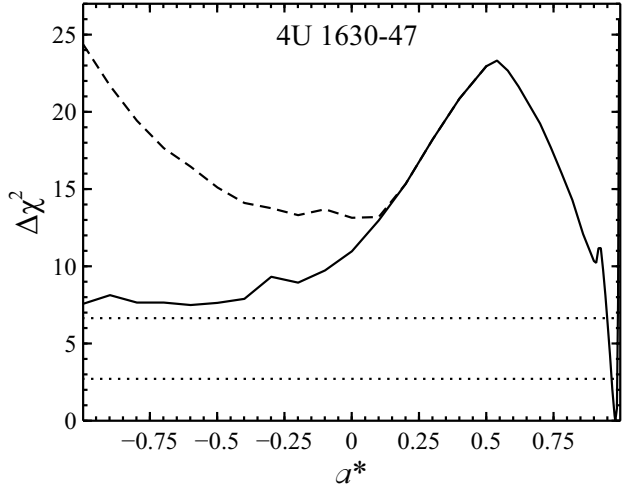} 
\caption{The goodness of fit in $\Delta\chi^2$ for the black hole spin, with the inner emissivity unconstrained (solid line), and the inner emissivity limited following \cite{Steiner12} below $a_{*}=0.5$ (dashed line). A high spin is clearly favored. The dotted lines show the 90\% (bottom) and 99\% (top) confidence levels. \label{fig:spin}}
\end{figure}

We show the change in the goodness of fit ($\Delta\chi^2$) for the spin with the {\tt refbhb} model in Figure \ref{fig:spin}. When the model is allowed sufficient freedom it appears as though there are low (retrograde) spin solutions that are also marginally acceptable. However, these solutions require very steep inner emissivity profiles, which are likely unphysical for such low spin values (\citealt{Wilkins11}). We therefore also show in Figure \ref{fig:spin} the change in the goodness of fit ($\Delta\chi^2$) obtained below $a_{*}=0.5$ when limiting the emissivity following \cite{Steiner12} (dashed line), demonstrating that low spin solutions are firmly rejected on physical grounds.

With previous radio ATCA and VLA observations described in \cite{Hjellming99} and our spin measurement, we can place 4U~1630$-$472 in the context of current spin-jet predictions. \cite{Narayan12} and \cite{Steiner13} report a correlation between the maximum outburst radio luminosity in X-ray binaries and spin: $P_{\rm{Jet}}\propto{a_*^2}$.  Assuming a distance of 10~kpc, mass of $10~M_\odot$, and the maximum observed radio flux density \citep[2.6mJy at 5 GHz,][]{Hjellming99}, the jet power as defined in \cite{Steiner13} is $P_{\rm jet}=1.5$~GHz~Jy~kpc~M$_\odot^{-1}$ (assuming a Doppler factor of 2). This puts 4U~1630$-$472 more than an order of magnitude below the relation predicted by \cite{Steiner13}, using our measured spin of $a_{*}=0.985$. Conversely, 4U~1630$-$472 would require a spin of $a_{*}\approx 0.25$ to agree with the relation given by \cite{Steiner13}. This is $4\sigma$ discrepant from our current measurements using the change in goodness of fit ($\Delta\chi^2=16$) for one degree of freedom (dof). A larger distance might reconcile the jet power prediction, but the X-ray luminosity would then exceed the Eddington limit for a 10~$M_\odot$ black hole. However, we note that the radio sampling was only every 3 days, so it is possible the true radio peak was missed.

\subsection{Absorption in a Disk Wind}
\label{sec:abs}
When the data are fit only with the reflection continuum with Galactic
absorption, the resulting chi-squared is $\chi^2/\nu$=1465/1091 (see panel 3 of Figure \ref{fig:del}). Including a Gaussian absorption line ({\it gabs}), we obtain an improvement of $\Delta\chi^{2}=229$ for 3 degrees of freedom
($\chi^2/\nu$=1255/1088). We find the line is detected at a 7.25$\sigma$ confidence level using an F-test. The line is centered at 7.03$\pm0.03$~keV, with a width of $\sigma=0.09\pm0.06$~keV~(3800$\pm2500$~km~s$^{-1}$)and an equivalent width
of 29$^{+5}_{-4}$~eV. Associating the line with Fe XXVI (6.97~keV), we
find a blue-shifted velocity of $v=0.0086\pm0.0043~c$~(2600$\pm1300$~km~s$^{-1}$). This is just
barely above the systematic centroiding uncertainty of {\it NuSTAR} (0.02~keV at this
energy), providing only tentative indication of an outflow. If instead we associate
the line with Fe XXV (6.70~keV), we find a blue-shifted velocity of
$v=0.049\pm0.005$~c~(14600$\pm1500$~km~s$^{-1}$).

In addition to a single Gaussian, we also tested a combination of two
narrow lines. The intrinsic energies of the two lines were fixed at
the rest frame energies of Fe XXV and FeXXVI, and we required them to
have a common outflow velocity. The line widths were also linked so
that both lines had the same relative broadening. However, two lines (Fe XXV and Fe XXVI) are not statistically favored over one line.

After the initial fits with a simple Gaussian, we utilized more
self-consistent, photoionization models with two XSTAR grids
\citep{Kallman01}. The initial grid was created assuming solar
metallicity, the spectral energy distribution derived from the
broadband continuum (unabsorbed powerlaw plus disk black body from 3--79~keV), and a constant gas density of
$n=10^{15}$cm$^{-3}$ \citep[X-ray wind densities range from
  $10^{13}$cm$^{-3}$ to 10$^{17}$cm$^{-3}$,
  e.g.,][]{Miller08,Miller13}. The ionizing luminosity was set at
$L_{\rm{ion}}$=10$^{38}$ergs s$^{-1}$, while the turbulent velocity
was set to 300~km~s$^{-1}$, which is typical of winds from other stellar-mass black holes. The second grid had the same parameters except the density was decreased to 10$^{12}$cm$^{-3}$ to explore the role of
density in the fits. The ionization, column density and velocity shift were all allowed to vary. 

\begin{figure*}
{\includegraphics[scale=0.5,angle=0]{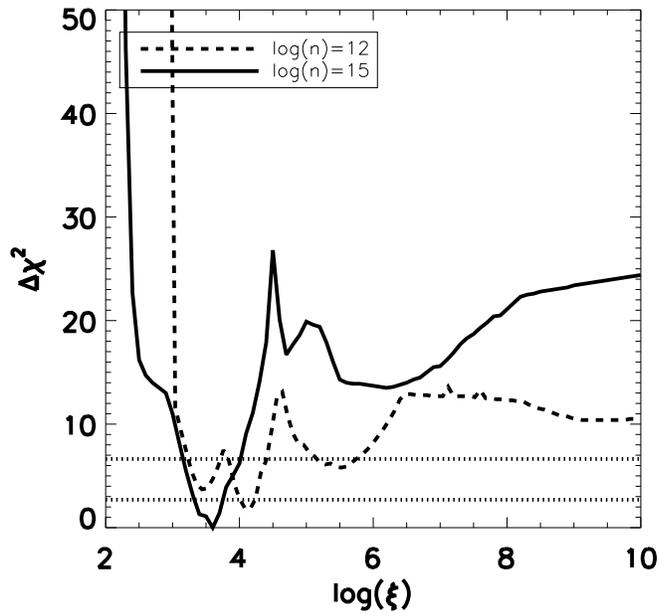}}  
\caption{The goodness of fit in $\Delta\chi^2$ for the ionization parameter of the photoionized absorber invoked for the absorption line at $\sim$7~keV. Both densities of 10$^{15}$cm$^{-3}$ (solid) and 10$^{12}$cm$^{-3}$ (dashed) show several minima corresponding to associating the 7~keV feature with different relative combinations of  Fe XXV and Fe XXVI. The dotted lines show the 90\% (bottom) and 99\% (top) confidence levels. \label{fig:cont_xi}}
\end{figure*}

The best fit model utilized the 10$^{15}$cm$^{-3}$ XSTAR grid
($\chi^2/\nu=1237/1088$), and the parameters obtained are given in
Table 2. As shown in Figure \ref{fig:cont_xi}, the ionization parameter, $\xi$, has
several local minima in the goodness of fit $\Delta\chi^2$ distribution. These minima are dependent on whether Fe XXV or Fe XXVI is predominantly associated with the 7 keV feature. This is determined by the strength of the optical depth of these lines in the XSTAR grids.  If the 7~keV feature is associated with Fe XXV, this corresponds to the lowest trough in Figure \ref{fig:cont_xi} ($\log\xi=3.6^{+0.2}_{-0.3}$ for n=10$^{15}$cm$^{-3}$), and an ultra-fast outflow, $v/c=0.043$~(12900~km~s$^{-1}$) is implied.  If instead the 7~keV feature is better associated with Fe XXVI, only a marginal blue-shift is required and the gas has an ionization of $\log\xi=6.1^{+0.7}_{-0.6}$. 

With a single broad feature and no direct density diagnostic, it is difficult to uniquely constrain all aspects of the wind. Using XSTAR, we investigated the effect of a much lower density, e.g. 10$^{12}$~cm$^{-3}$ \citep[as assumed in a wind component by][]{Neilsen12}.  As expected, a gas with a lower density will be more highly ionized, but will still be dominated by Fe XXV for all of the troughs seen in Figure \ref{fig:cont_xi}. Conversely, a higher density \citep{Kallman01} or increase in Fe abundance \citep{King12} would lower the ionization for a given strength of a particular line species.

Previous work by \cite{Kubota07} find resolved Fe XXV and Fe XXVI at 6.7~keV and 7~keV, respectively, at
approximately an ionization of $\log\xi\approx4.6$. However, the \cite{Kubota07} observations indicate a lower velocity shift and have a lower ionizing flux.  Our higher ionizing flux hints to the high ionization models being a more physical interpretation of our data. This is also supported by the fact that the tentative velocity implied by the high ionization ($\log\xi=6.1^{+0.7}_{-0.6}$) is closer to the line shifts observed by \cite{Kubota07}.

We can roughly estimate the mass outflow rate if we assume a modified spherical outflow: $\dot{M}_{\rm{wind}}=\Omega\mu m_p nr^2v=\Omega \mu m_p L_{\rm{ion}} v /\xi$, where $\Omega$ is the opening angle (assumed to be 2$\pi$), $\mu$ is the average weight of the gas, $m_p$ is the mass of a proton, and we used $\xi=L_{\rm{ion}}/nr^2$ to convert the radius and number density to direct observable quantities ($L_{\rm{ion}},~\xi$). We find that the mass outflow rate is 5$\times10^{20}g~s^{-1}$ ($\dot{M}_{\rm{wind}}/\dot{M}_{\rm{acc}}=400$) for an ionization of $\log\xi=3.6^{+0.2}_{-0.3}$ and $5\times10^{17}g~s^{-1}$ ($\dot{M}_{\rm{wind}}/\dot{M}_{\rm{acc}}=0.4$) for an ionization of $\log\xi=6.1^{+0.7}_{-0.6}$. \cite{Ponti12} show in their figure 4 a relation between wind outflow rates and Eddington fraction. Neither of our mass outflow rate estimates fall on this relation, though we can not rule out an intermediate ionization and velocity that would bring the mass outflow rate closer to the relation given by \cite{Ponti12}. The only other outlier in this relation is GRO~J1655$-$40, which is likely magnetically driven \citep{Miller08,Neilsen12}, while the other winds are thought to be Compton heated. Though the mass accretion rate of 4U~1630$-$472 may not agree with other BHB, its high inclination is consistent with BHB X-ray winds being equatorial \citep[e.g.,][]{Miller08,Ponti12}.

\section{Discussion}
In this Letter, we have analyzed an early 14.7 ks {\it NuSTAR}
observation of 4U~1630$-$472 to determine the key properties of this
well-known accreting X-ray binary.  We have found this
source to be in a intermediate state dominated primarily by the
accretion disk, but also displaying a hard Comptonized tail and clear
reflection features from the inner disk. This is the first time such features have unambiguously been seen in this source. The resulting spin measurement indicates that the
black hole is rapidly spinning: $a_{*}=0.985^{+0.005}_{-0.014}$ (1$\sigma$ statistical errors).  In
addition, we found clear evidence for ionized absorption at 7.03 keV.
Through photoionization modeling with XSTAR, we find that the
absorbing material is highly ionized.

We cannot unambiguously determine the nature of this outflow, given a
single spectral feature and modest spectral resolution, but the
possibilities are extremely interesting.  The outflow may represent an
ultra-fast outflow.  If this is correct, the wind could be even more
extreme than that detected in the black hole candidate IGR
J17091$-$3624 \citep{King12}, and may be analogous to the flows detected
in e.g. PDS 456 \citep{Reeves03}.
Alternatively, the disk wind may have a much lower velocity but
originate very close to the black hole.  The outflow may originate as
close as $200~R_G<R<1100~R_G$ based on ionization, suggesting that the disk wind is arising
from the inner accretion disk.  In X-ray binaries, winds can be
launched either by radiative, thermal or magnetic processes.  At such
high ionizations, line driving becomes inefficient \citep{Proga00},
making it an unlikely mechanism.  In addition, thermal driving
is only efficient in the outer accretion disk \citep{Woods96}.
Therefore, a magnetically driven wind is the most likely launching
mechanism \citep[e.g,][]{Blandford82}, as also suggested by \cite{Kubota07} and similar to GRO~J1655$-$40
\citep{Miller08}.

Ionized absorption has also been previously observed in 4U 1630$-$472. In 2000, \cite{Cui00} previously reported the presence of 2 Fe K lines based on {\it RXTE} data. However, in view of our results, the putative Doppler-split line pair is probably best understood in terms of a single relativistic emission line, with absorption features superimposed.  Whereas {\it RXTE} lacked the resolution to accurately decompose the observed spectrum, this situation is very clear with {\it NuSTAR}.

More recently, \cite{Kubota07} detected two absorption features at 6.73 keV and 7.0
keV in six {\it Suzaku} observations of 4U 1630$-$472 taken between 2006
February 8 and March 23. During their
six observations the ionization decreases as indicated by the decreasing ratio of
line strength between the Fe XXV and Fe XXVI. Our observation either suggests a higher ionization at similar outflow velocity, or a lower ionization but with a higher outflow velocity.

Finally, if our spin measurement and current proxies for jet power are
both correct, the radio emission from 4U~1630$-$472 is too weak to
agree with current spin-jet relations reported by \cite{Steiner13}. It is likely that other aspects of the accretion flow are also responsible for the ultimate magnitude of the jet power, as suggested
by \cite{King13b} and \cite{Tomsick05}.  Observations across other wavelengths, including
radio, will be needed to follow the evolution of this source and its
outflows in the future.

\begin{acknowledgements}
We would like to thank Julia Lee for her invaluable comments. This work was supported under NASA Contract No. NNG08FD60C, and made use of data from the {\it NuSTAR} mission, a project led by the California Institute of Technology, managed by the Jet Propulsion Laboratory, and funded by NASA. LN wishes to acknowledge the Italian Space Agency (ASI) for financial support
by ASI/INAF grant I/037/12/0-011/13"
 \end{acknowledgements}

\begin{deluxetable*}{c c c c c c c}[h]
\centering
\tablecolumns{6}
\tablewidth{0pc}
\tabletypesize{\scriptsize}
\tablecaption{Initial continuum parameters}
\tablehead{ $\Gamma$ & powerlaw norm & kT$_{\rm BB}$ & DiskBB norm & $N_{\rm H}$ & FPMA/FPMB & $\chi^2/\nu$ \\
& (photons keV$^{-1}$cm$^{-2}$s$^{-1}$) &  (keV) & & (10$^{22}$cm$^{-2}$) & cross norm & }
\startdata
2.09$\pm$0.04 & 0.20$^{+0.03}_{-0.02}$& 1.450$\pm$0.003 & 198$\pm$3 & 8.09$\pm0.07$&  0.990$\pm$0.002 & 1945.13/1097 \\

\enddata  
\label{tab:corr}
\tablecomments{\small{The $\chi^2/\nu$=1945.13/1097=1.77.  There are clear indications of a broad Fe K$\alpha$ line with Compton reflection at high energies, as well as a strong absorption feature at approximately 7~keV. See Figure \ref{fig:del}.} }

\end{deluxetable*} 

\begin{deluxetable*}{l l l }[h]
\centering
\tablewidth{0pc}
\tabletypesize{\scriptsize}
\tablecaption{Parameters for the best fit model for n=10$^{15}$~cm$^{-3}$}

\tablehead{Model & $\log\xi=3.6$ &  $\log\xi=6.1$  \\ Parameters}
\startdata
N$_{\rm{H}}$ (10$^{22}{\rm{cm}}^{-2}$) & 8.0$_{-0.3}^{+0.2}$ & 8.1$^{+0.2}_{-0.3} $ \\
FPMA/FPMB & 0.990$\pm0.002$  &  0.990$\pm0.002$ \\
$\Gamma$ & 1.92$_{-0.15}^{+0.03}$  &  1.92$_{-0.06}^{+0.04}$ \\ 
kT$_{\rm BB}$ (keV) & 1.02$\pm0.01$ &  1.02$^{+0.01}_{-0.02} $\\ 
nthComp Norm ($\times10^{-3}$) & 2.0$^{+3.5}_{-2.0}$ & 2.0$^{+1.4}_{-2.0}$ \\
$q_{\rm{in}} $&10.0$^{~\diamond}_{-0.7}$  &10.0$^{~\diamond}_{-0.4}$ \\
 $q_{\rm{out}}$ & 3.0*  &3.0*  \\
R$_{\rm Br}$  (R$_{\rm G}$) & 2.7$_{-0.1}^{+0.2}$ &  2.7$^{+0.2}_{-0.1}$ \\
 $a_*$ & 0.985$_{-0.014}^{+0.005}$  & 0.985$^{+0.003}_{-0.010}$  \\
$\theta$ ($^\circ$) & 64$^{+2}_{-3}$ & 64$\pm2$  \\
H$_{\rm Den}$ (10$^{21}$cm$^{-3}$) & 4.8$^{+0.7}_{-1.3}$ &  4.7$_{-1.2}^{+0.8}$ \\
 Illum/BB  & 0.18$_{-0.08}^{+0.01}$ &   0.18$\pm0.05$ \\
 $\log\xi_{\rm REFBHB}$ & 2.9$\pm$0.2 & 2.9$\pm$0.3 \\
REFBHB Norm & 0.15$_{-0.09}^{+0.01}$ &  0.15$\pm0.03$\\
$N_{\rm{H;xstar}}$ (10$^{22}{\rm{cm}}^{-2}$)&2.4$_{-0.9}^{+0.7}$  & 93$^{+7~\diamond}_{-45}$  \\
$\log\xi_{\rm{xstar}}$ & 3.6$^{+0.2}_{-0.3}$ &   6.1$^{+0.7}_{-0.6}$\\
$z_{\rm{xstar}}$ & 0.043$^{+0.002}_{-0.007}$ &  0.014$^{+0.002}_{-0.005}$  \\
v$_{\rm{xstar}}$ (km~s$^{-1}$ ) & 12900$^{+600}_{-2100}$ & 4200$^{+600}_{-1500}$\\
$\chi^2/\nu$ & 1237/1088  & 1250/1088 \\

\enddata  
\label{tab:corr}
\tablecomments{ These are the best fit parameters when including both reflection and the XSTAR photoionization model (v$_{\rm{turbulence}}$=300 km s$^{-1}$, $n=10^{15}$~cm$^{-3}$). The best fit parameters for the three lowest minima shown in Figure \ref{fig:cont_xi} are shown here. $\diamond$ reached the upper limit. $*$ frozen}

\end{deluxetable*} 


\end{document}